\documentclass{article}
\usepackage{subcaption}
\usepackage{spconf,amsmath,amssymb}

\usepackage{hyperref}
\usepackage{url}
\usepackage{booktabs}       % professional-quality tables
\usepackage{amsfonts}       % blackboard math symbols
\usepackage{nicefrac}       % compact symbols for 1/2, etc.
\usepackage{microtype}      % microtypography
\usepackage{xcolor}         % colors
\global\definecolor{darkgray}{gray}{0.3}
\definecolor{ired}{RGB}{255,59,48}
\global\definecolor{iorange}{RGB}{255,149,0}
\definecolor{iyellow}{RGB}{255,204,10}
\definecolor{igreen}{RGB}{52,199,89}
\definecolor{iblue}{RGB}{0,122,255}
\definecolor{ipurple}{RGB}{175,82,222}
\usepackage{xspace}    
\usepackage{tabularx}         
\usepackage{amstext} % for \text macro
\usepackage{array}   % for \newcolumntype macro
\usepackage{algorithm,algorithmicx,algpseudocode}

\usepackage[pdftex]{graphicx}

\usepackage{tikz}

\usepackage{pgfplots, pgfplotstable}
\pgfplotsset{compat=1.17}
\usetikzlibrary{calc, shapes}
\usepackage{amsmath,amsfonts,bm}

\def\eqref#1{equation~\ref{#1}}
\def\1{\bm{1}}

\def\vmu{{\bm{\mu}}}
\def\vphi{{\bm{\phi}}}

\def\vx{{\bm{x}}}

\def\mI{{\bm{I}}}

\def\ms{{\bm{s}}}

\def\mX{{\bm{X}}}

\DeclareMathAlphabet{\mathsfit}{\encodingdefault}{\sfdefault}{m}{sl}
\SetMathAlphabet{\mathsfit}{bold}{\encodingdefault}{\sfdefault}{bx}{n}

\newcommand{\tableresults}{
\newcolumntype{C}{>{$}c<{$}} % math-mode version of "l" column type
\begin{table*}[th]
  \caption{The evaluation results for models with and without PhaseAug. The bold numbers indicate the better value between with and without PhaseAug. Paired bold numbers indicate that standard deviation or confidence interval ranges are overlapped.}
  \label{tab:result}
  \vspace{-0.125\baselineskip}
  \centering
  \resizebox{\linewidth}{!}{%
  \begin{tabular}{l |CCCCCC |CC}
    \toprule
    Model & \text{MAE } (\downarrow) & \text{M-STFT } (\downarrow) & \text{PESQ } (\uparrow) & \text{MCD } (\downarrow) & \text{V/UV F$1$ } (\uparrow) &     \text{Periodicity } (\downarrow) & \text{MOS } (\uparrow) & \text{Pairwise } (\uparrow) \\
    \midrule
    Ground Truth &-&-&-&-&-&-& \bm{4.08} \pm 0.05 & - \\
    + PhaseAug      &0.02368 & 0.2585 & 4.608 & 0.1740 & 0.9949 & 0.02062 & \bm{4.03} \pm 0.06  & -\\
    \midrule 
    HiFi-GAN ($100\%$)     
    &0.2227    & {1.004} & 3.634 & \bm{1.012}  & 0.9582 & {0.1111} & \bm{3.91}\pm 0.06  & 32.9\%\\ %2.5M
    + PhaseAug  & \bm{0.2111} & \bm{1.001} & \bm{3.667} & \bm{1.012} & \bm{0.9603} & \bm{0.1053} & \bm{3.99} \pm0.05  & \bm{38.1\%}\\
        \specialrule{0.01em}{0.01em}{0.01em}
    MelGAN ($100\%$)           & 0.2849 & 1.115 & 2.924 & 1.154 & 0.9477 & 0.1325 & \bm{3.98} \pm 0.06  & 32.9\%\\
    + PhaseAug  & \bm{0.2651} & \bm{1.091} & \bm{3.037} & \bm{1.081}  & \bm{0.9503} & \bm{0.1270} & \bm{3.96}\pm0.06  & \bm{39.7\%}\\%var sqrt6

    \midrule
    HiFi-GAN ($10\%$)          & 0.2395 & 1.033 & 3.494 & 1.104 & 0.9543 & 0.1169 & \bm{3.97}\pm 0.06  & 29.2\%\\%2.35M
    + PhaseAug  & \bm{0.2202} & \bm{1.012} & \bm{3.600} & \bm{1.095} & \bm{0.9568} & \bm{0.1129} & \bm{3.96} \pm 0.06  & \bm{40.6\%} \\%sqrt 6 2.35M
        \specialrule{0.01em}{0.01em}{0.01em} 
    MelGAN ($10\%$)           & 0.2967 & 1.121 & 2.860 & \bm{1.177} & 0.9443 & 0.1399 & \bm{3.92} \pm0.06  & 31.1\%\\
    + PhaseAug    & \bm{0.2824} & \bm{1.117} & \bm{2.891} & 1.271 & \bm{0.9450} & \bm{0.1362} & \bm{3.98} \pm0.06  & \bm{43.9\%}\\%var sqrt6
    \midrule
   HiFi-GAN ($1\%$)          & 0.3383 \footnotesize{\pm 0.003} & 1.177 \footnotesize{\pm 0.003} & \bm{2.703} \footnotesize{\pm 0.02} & \bm{1.320} \footnotesize{\pm 0.01} & \bm{0.9350} \footnotesize{\pm 0.001} & 0.1576 \footnotesize{\pm 0.002} & \bm{3.80} \pm 0.05  & 33.5\%\\
    + PhaseAug  & \bm{0.3302} \footnotesize{\pm 0.004}  & \bm{1.169} \footnotesize{\pm 0.003} & \bm{2.726 } \footnotesize{\pm 0.04} & \bm{1.314} \footnotesize{ \pm 0.05} & \bm{0.9363} \footnotesize{ \pm 0.001} & \bm{0.1543} \footnotesize{ \pm 0.001} & \bm{3.88} \pm 0.06   & \bm{43.3\%}\\ %mean std
        \specialrule{0.01em}{0.01em}{0.01em} 

    MelGAN ($1\%$)             & 0.3906 \footnotesize{\pm 0.004} & 1.275 \footnotesize{\pm 0.002} & 2.020 \footnotesize{\pm 0.005} & \bm{1.438} \footnotesize{ \pm 0.02} & \bm{0.9256} \footnotesize{ \pm 0.001} & \bm{0.1811} \footnotesize{ \pm 0.002} & \bm{3.65} \pm 0.07  & \bm{36.4\%}\\%97k    
    + PhaseAug   & \bm{0.3756}\footnotesize{ \pm 0.004} & \bm{1.257}\footnotesize{ \pm 0.002} & \bm{2.093}\footnotesize{ \pm 0.02} & \bm{1.416} \footnotesize{ \pm 0.05} & \bm{0.9248} \footnotesize{ \pm 0.002} & \bm{0.1803} \footnotesize{ \pm 0.002} & \bm{3.56} \pm 0.07  & \bm{38.5\%}\\%96k sqrt6
    \bottomrule
  \end{tabular}
}
\vspace{-0.675\baselineskip}
\end{table*}
}

\usepackage{enumitem}
\let\Phi\varPhi

\makeatletter
\DeclareRobustCommand\onedot{\futurelet\@let@token\@onedot}
\def\@onedot{\ifx\@let@token.\else.\null\fi\xspace}

\def\etc{\emph{etc}\onedot} 
 
\def\etal{\emph{et al}\onedot}
\makeatother
\title{PhaseAug: A Differentiable Augmentation for Speech Synthesis\\to Simulate One-to-Many Mapping}
\name{Junhyeok Lee$^{1}$\thanks{This work was supported by Institute for Information \& communications Technology Planning \& Evaluation(IITP) grant funded by the Korea government(MSTI) (No. 2021-0-00062-002, Development of joint work automation management software technology based on task awareness)}
\qquad Seungu Han$^{2,4\ast}$\thanks{$\ast$ Work performed at MINDsLab Inc.} \qquad Hyunjae Cho$^{1,2}$ \qquad Wonbin Jung$^{1,3}$}
\address{$^{1}$ MINDsLab Inc., Republic of Korea \\$^{2}$ Seoul National University (SNU), Republic of Korea \\$^{3}$ Korea Advanced Institute of Science and Technology (KAIST), Republic of Korea,\\$^{4}$ Supertone Inc., Republic of Korea}

\begin{document}
%\ninept

% \SetWatermarkText{Preprint MINDsLab Inc. jun3518 Preprint MINDsLab Inc. jun3518 Preprint MINDsLab Inc. jun3518
% \\MINDsLab Inc. jun3518 Preprint MINDsLab Inc. jun3518 Preprint MINDsLab Inc. jun3518
% } % Text to be printed across the page
% \SetWatermarkScale{1} % Size of the watermark text
%\SetWatermarkColor[gray]{0.95}

\maketitle
\begin{abstract}
Previous generative adversarial network (GAN)-based neural vocoders are trained to reconstruct the exact ground truth waveform from the paired mel-spectrogram and do not consider the one-to-many relationship of speech synthesis. This conventional training causes overfitting for both the discriminators and the generator, leading to the periodicity artifacts in the generated audio signal. In this work, we present PhaseAug, the first differentiable augmentation for speech synthesis that rotates the phase of each frequency bin to simulate one-to-many mapping. With our proposed method, we outperform baselines without any architecture modification. Code and audio samples will be available at \url{https://github.com/mindslab-ai/phaseaug}.
\end{abstract}
\begin{keywords}
speech synthesis, neural vocoder, differentiable augmentation, phase, generative adversarial networks
\end{keywords}
\vspace{-0.64\baselineskip}
\section{Introduction}
\vspace{-0.64\baselineskip}
Neural vocoder synthesizes raw audio waveform from input acoustic feature, such as mel-spectrogram or latent variable. 
While a variety of generative models are applied to vocoder task, generative adversarial networks (GANs) \cite{gan} are the most commonly used generative models for vocoder task \cite{melgan,vocgan,hifigan,fregan,cargan,bigvgan,avocodo}.
GAN-based vocoder models after MelGAN \cite{melgan} are mainly focused on modifying GAN architectures,
among these, HiFi-GAN \cite{hifigan} utilizes two kinds of raw waveform discriminators, multi-scale discriminators (MSD) and multi-period discriminators (MPD), to achieve a state-of-the-art GAN-based non-autoregressive vocoder.
However, during training HiFi-GAN, we observed training accuracies of MSD and MPD converged to $100\%$ and over $70\%$ respectively, which indicates discriminator overfitting.

For GAN-based image synthesis, discriminator training methods are being actively studied, such as differentiable augmentation (DiffAugment) \cite{diffaug} and adaptive discriminator augmentation (ADA) \cite{stylegan2ada}.
These methods apply DiffAugment to real data and generated output to prevent overfitting of the discriminator and to provide useful gradients for the generator.
While there are speech synthesis studies adapting data augmentations \cite{ttsbytts, nansy, mlttsvcsl}, there are no speech synthesis studies adapting DiffAugment to GAN to the best of our knowledge.

GAN-based non-autoregressive vocoders also suffer from a one-to-one paired data problem.
While neural vocoding is a one-to-many mapping problem, previous neural vocoders were trained assuming a one-to-one mapping between the ground truth waveforms and the paired mel-spectrograms.
Recently, Morrison \etal{} \cite{cargan} claim that non-autoregressive generators of vocoders are forced to generate the exact phases of ground truth signal from the mel-spectrogram without providing the initial phase.
In addition, Morrison \etal{} \cite{cargan} report that HiFi-GAN generates the perceptible periodicity artifact and suggest periodicity error, which is extracted from the pitch estimation model CREPE \cite{crepe}, to measure it.
We hypothesize that the periodicity artifact is related to a one-to-one paired data training, since generators are overfitted to generate a given phase of each ground truth waveform.

%reviwer comment: artifact 없이 augment 한게 contribution인데 좀 더 강조 어떠냐
To overcome the discriminator overfitting and the one-to-one paired data problem, we propose \textit{PhaseAug}, the first differentiable augmentation for audio signal synthesis, to simulate one-to-many mapping by rotating signal's phases on the frequency domain.
Compared to conventional trained HiFi-GAN and MelGAN, we outperform each of the models by simply applying PhaseAug to these models.

\vspace{-0.64\baselineskip}
\section{PhaseAug}

\begin{figure*}[th!]
  \centering
  \includegraphics[width=0.99015\textwidth]{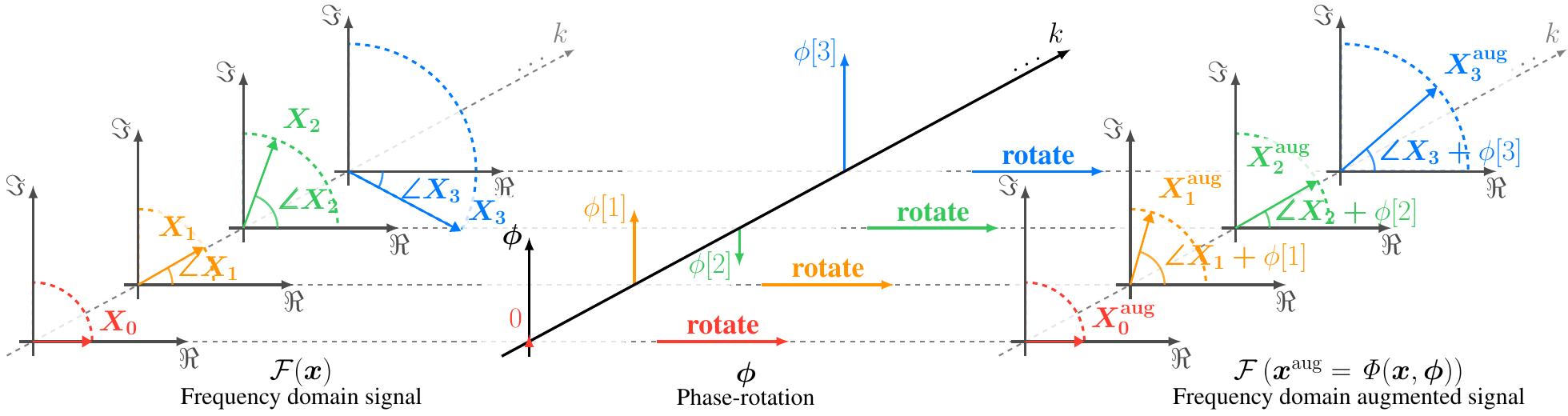}
      \vspace{-1.2\baselineskip}
\caption{
Our suggested differentiable augmentation operation, PhaseAug $\Phi$. We rotate signal $\vx$'s phase for each frequency bin as $\vphi$ to generate augmented signal $\vx^\mathrm{aug}$. $\mX_k$ for $0 \le k \le N/2$ are abbreviation of $\mX[k,m]$ for arbitrary $m$th time frame.
}
\label{fig:fig}
\end{figure*}

\vspace{-0.64\baselineskip}
\subsection{Fourier Transformation}
\vspace{-0.64\baselineskip}
In the signal processing domain, Fourier transformation is widely applied to analyze time-series signals in the frequency domain.
For a real-valued discrete-time digital signal $\vx=\vx [n]=\begin{bmatrix} \dotsc & x[0] & x[1] & x[2]&\dotsc \end{bmatrix}$ for $n \in \mathbb{Z}$, short-time Fourier transform (STFT) $\mathcal{F}$ and inverse short-time Fourier transform (iSTFT) $\mathcal{F}^{-1}$ are represented as:
\begin{equation}
    \mX[k,m] =  \mathcal{F}(\vx) = \sum_{n=-\infty}^{\infty} x[n] w[n-md] e^{-j 2 \pi nk /N},
\end{equation}
\begin{equation}
    \vx[n]= \mathcal{F}^{-1}(\mX[k,m])  =\frac{1}{C} \sum_{k=0}^{N-1}  \sum_{m=-\infty}^{\infty}  \mX[k,m] e^{j 2 \pi nk/N},
\end{equation}
where $N$ is the size of STFT, $d$ is the hop size, $[k, m]$ are the respective frequency bin and time frame indices of Fourier feature $\mX$, $w[n]$ is the Hann-window $w[n]=(1-\cos(2 \pi n/N))/2$ for $0\leq n\leq N-1$ else $0$, and $C=\sum_{m=-\infty}^{\infty}w[n-md]$ is the overlap-add value of Hann-window. 
Since real-valued $\vx$ satisfies $\mX[k,m]=\mX[N-k,m]^\ast,\, \forall k \in \mathbb{Z}$, we can consider a one-sided STFT in $0 \leq k \leq N/2$ as $\mX[k,m] \in \mathbb{C}^{(N/2+1) \times M}$.
We start all vector and matrix indices at $0$, since the range of $k$ starts at $0$.
From the definition, STFT of $\vx [n-\delta]$, for a small enough time-shift $\delta \ll N$ on signal $\vx[n]$, can be approximated as:
\begin{align} \label{eq_time_shift}
    \mathcal{F}(\vx[n-\delta]) & = \sum_{n=-\infty}^{\infty} x[n-\delta] w[n-md] e^{-j 2 \pi nk /N} \nonumber \\ 
    & \approx  \mX[k,m] e^{-j 2 \pi \delta k  /N},
\end{align}
since difference of window is negligible for small enough $\delta$.

\subsection{PhaseAug}
The one-to-many relationship between mel-spectrogram and raw waveform is not considered by previous studies, thus vocoders were forced to generate exact ground truth waveforms.
We present \textit{PhaseAug} which arbitrarily rotates the phase of each frequency bin without manipulating magnitude to simulate one-to-many mapping from paired data.
We define PhaseAug operation $\Phi$ as:
\begin{equation} \label{eq_phaseaug}
    \Phi\left(\vx,\vphi \right) = \mathcal{F}^{-1}\left(\mathrm{\textbf{diag}} \left(e^{j \vphi}\right)\mathcal{F}\left(\vx\right)\right),
\end{equation} 
where $\vphi = \begin{bmatrix} 0 & \phi[1] & \phi[2]& \dotsc & \phi[N/2] \end{bmatrix}\in \mathbb{R}^{(N/2+1)}$ is a vector of phase-rotation for each frequency bin in radian, and $\mathrm{\textbf{diag}}:\mathbb{C}^{(N/2+1)} \rightarrow \mathbb{C}^{(N/2+1)\times (N/2+1)}$ denotes vector-to-matrix operator that maps elements of the vector to the main diagonal elements of a square diagonal matrix.
From (\ref{eq_time_shift}), PhaseAug operation is identical to rotating the phase of $k$th frequency bin as $\phi [k]$ without modifying the magnitude and the relative phase between the adjacent time frames.
We always set $\phi[0]=0$ to prevent DC offset and complex output.
Since all of the components, including STFT and iSTFT, are differentiable,
PhaseAug is applicable as DiffAugment \cite{diffaug}.
We expect that PhaseAug could reduce the periodicity artifacts by preventing the discriminators and the generator from overfitting to the ground truth phase.

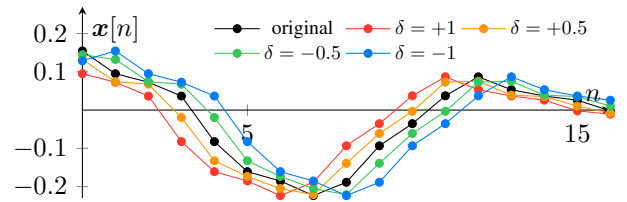
\begin{figure}[h]
  \centering
  \begin{tikzpicture}
    \usepgfplotslibrary{polar}
    \usepgflibrary{shapes.geometric}
    \pgfplotsset{mystyle/.append style={axis x line=middle,
    axis y line=
           middle,
    xlabel={$n$}, ylabel={$\vx[n]$},
    xticklabel={$\pgfmathprintnumber{\tick}$},
    yticklabel={$\pgfmathprintnumber{\tick}$}}
    }

    \begin{axis}[mystyle,width = \linewidth,height = 0.48\linewidth,
    legend style={draw=none, fill opacity=1., text opacity = 1,row sep=0pt,nodes={scale=0.75, transform shape}},
    legend columns=3,% transpose legend,    
    %legend pos =outer north west,
    xtick={0,5,15},%xtick={0,5,...,50},
    ytick={-0.2,-0.1,0,0.1,0.2}, xmin=0, xmax=16, ymin=-0.23, ymax=0.27]
    %\addlegend{original,$\delta=-1$,$\delta=-0.5$,$\delta=+0.5$,$\delta=+1$}
        \addplot[mark=*, color = black,mark options  = {scale = 0.75}] coordinates {(0,  0.154388427734375)(1,  0.095062255859375)(2,  0.07275390625)(3,  0.036590576171875)(4,  -0.082427978515625)(5,  -0.160919189453125)(6,  -0.185302734375)(7,  -0.2235107421875)(8,  -0.1885986328125)(9,  -0.093353271484375)(10,  -0.035491943359375)(11,  0.037322998046875)(12,  0.086883544921875)(13, 0.053131103515625)(14,  0.036712646484375)(15,  0.02581787109375)(16,  -0.00213623046875)(17, -0.01080322265625)(18,  -0.025634765625)(19,  -0.053985595703125)(20,  -0.0794677734375)(21, -0.096893310546875)(22,  -0.103851318359375)(23,  -0.091949462890625)(24,  -0.07049560546875)(25, -0.085723876953125)(26,  -0.147857666015625)(27,  -0.215118408203125)(28,  -0.265228271484375)(29, -0.281982421875)(30,  -0.258453369140625)(31,  -0.22052001953125)(32,  -0.208099365234375)(33, -0.18731689453125)(34,  -0.1815185546875)(35,  -0.18670654296875)(36,  -0.176513671875)(37, -0.161529541015625)(38,  -0.147918701171875)(39,  -0.121826171875)(40,  -0.11328125)(41,  -0.118408203125)(42,  -0.1231689453125)(43,  -0.11834716796875)(44,  -0.093780517578125)(45, -0.053680419921875)(46,  -0.029083251953125)(47,  -0.01934814453125)(48,  -0.041595458984375)(49,  -0.100860595703125)}; \addlegendentry{original};
        \addplot[mark=*, color = ired, mark options  = {scale = 0.75}] coordinates {(0,  0.09506163746118546)(1,  0.07275346666574478)(2,  0.036590319126844406)(3,  -0.08242745697498322)(4,  -0.1609182059764862)(5,  -0.1853015422821045)(6,  -0.22350937128067017)(7,  -0.18859745562076569)(8,  -0.09335268288850784)(9,  -0.03549171984195709)(10,  0.037322770804166794)(11,  0.08688300848007202)(12, 0.053130753338336945)(13,  0.03671243041753769)(14,  0.02581770531833172)(15, -0.002136208349838853)(16,  -0.010803145356476307)(17,  -0.025634611025452614)(18, -0.05398526415228844)(19,  -0.07946724444627762)(20,  -0.09689267724752426)(21, -0.10385064035654068)(22,  -0.09194886684417725)(23,  -0.07049515098333359)(24, -0.08572331815958023)(25,  -0.14785674214363098)(26,  -0.21511709690093994)(27, -0.2652266025543213)(28,  -0.28198060393333435)(29,  -0.2584517300128937)(30, -0.22051863372325897)(31,  -0.20809805393218994)(32,  -0.18731571733951569)(33, -0.18151742219924927)(34,  -0.1867053508758545)(35,  -0.17651256918907166)(36, -0.16152849793434143)(37,  -0.14791779220104218)(38,  -0.12182542681694031)(39, -0.11328054219484329)(40,  -0.1184074878692627)(41,  -0.12316814810037613)(42, -0.11834641546010971)(43,  -0.09377994388341904)(44,  -0.053680066019296646)(45, -0.02908307872712612)(46,  -0.019348019734025)(47,  -0.041595205664634705)(48, -0.10085994005203247)(49,  0.018676670268177986)}; \addlegendentry{$\delta=+1$};
        \addplot[mark=*, color = iorange, mark options  = {scale = 0.75}] coordinates {(0,  0.13218456506729126)(1,  0.07389122992753983)(2,  0.06762560456991196)(3,  -0.019511837512254715)(4,  -0.132571280002594)(5,  -0.17343974113464355)(6,  -0.2051514983177185)(7,  -0.22084327042102814)(8,  -0.1391194462776184)(9,  -0.06121879443526268)(10,  -0.0033640952315181494)(11,  0.07314108312129974)(12, 0.07520929723978043)(13,  0.03901613876223564)(14,  0.03546806052327156)(15,  0.010343766771256924)(16,  -0.007933385670185089)(17,  -0.015980349853634834)(18, -0.039035167545080185)(19,  -0.06793250888586044)(20,  -0.08894668519496918)(21, -0.10239889472723007)(22,  -0.10035622119903564)(23,  -0.08027061074972153)(24, -0.07058248668909073)(25,  -0.11378052085638046)(26,  -0.18244881927967072)(27, -0.24366910755634308)(28,  -0.27816784381866455)(29,  -0.275403767824173)(30, -0.23697568476200104)(31,  -0.21288444101810455)(32,  -0.19904768466949463)(33, -0.18047688901424408)(34,  -0.18559934198856354)(35,  -0.1831880360841751)(36, -0.16864967346191406)(37,  -0.1555715948343277)(38,  -0.13569633662700653)(39, -0.11347129940986633)(40,  -0.1164655014872551)(41,  -0.12014897912740707)(42,  -0.123921237885952)(43,  -0.10753979533910751)(44,  -0.07563921064138412)(45, -0.036059826612472534)(46,  -0.026260368525981903)(47,  -0.01817161776125431)(48, -0.08276288956403732)(49,  -0.062493544071912766)}; \addlegendentry{$\delta=+0.5$};
        \addplot[mark=*, color = igreen, mark options  = {scale = 0.75}] coordinates {
        (0,  0.14452780783176422)(1,  0.13218456506729126)(2,  0.07389122247695923)(3,  0.06762559711933136)(4,-0.019511835649609566)(5,  -0.1325712949037552)(6,  -0.17343972623348236)(7,-0.20515145361423492)(8,  -0.22084330022335052)(9,  -0.1391194611787796)(10,  -0.06121878698468208)(11,  -0.0033641087356954813)(12, 0.07314109057188034)(13,  0.07520926743745804)(14,  0.03901612386107445)(15,  0.03546804562211037)(16,  0.010343757458031178)(17,  -0.007933381013572216)(18,  -0.015980353578925133)(19, -0.039035167545080185)(20,  -0.06793248653411865)(21,  -0.08894669264554977)(22, -0.10239890217781067)(23,  -0.10035621374845505)(24,  -0.08027059584856033)(25, -0.07058247178792953)(26,  -0.11378052830696106)(27,  -0.18244880437850952)(28, -0.24366913735866547)(29,  -0.27816781401634216)(30,  -0.27540382742881775)(31, -0.23697571456432343)(32,  -0.21288445591926575)(33,  -0.19904766976833344)(34, -0.18047690391540527)(35,  -0.18559932708740234)(36,  -0.1831880807876587)(37, -0.16864968836307526)(38,  -0.1555716097354889)(39,  -0.13569632172584534)(40, -0.11347132176160812)(41,  -0.1164654865860939)(42,  -0.12014901638031006)(43,  -0.1239212155342102)(44,  -0.1075398325920105)(45,  -0.07563921064138412)(46, -0.036059826612472534)(47,  -0.026260359212756157)(48,  -0.01817162334918976)(49, -0.08276285976171494)
        }; \addlegendentry{$\delta=-0.5$};
        \addplot[mark=*, color = iblue, mark options  = {scale = 0.75}] coordinates {(0,  0.12860025465488434)(1,  0.1543874889612198)(2,  0.09506163001060486)(3,  0.07275346666574478)(4,  0.03659031167626381)(5, -0.08242744952440262)(6,  -0.1609182208776474)(7,  -0.1853015273809433)(8,  -0.22350935637950897)(9, -0.1885974407196045)(10,  -0.09335270524024963)(11,  -0.035491716116666794)(12,  0.037322748452425)(13, 0.08688299357891083)(14,  0.05313075706362724)(15,  0.03671242669224739)(16,  0.02581769973039627)(17, -0.0021362188272178173)(18,  -0.010803140699863434)(19,  -0.025634588673710823)(20,  -0.05398525297641754)(21,  -0.07946725934743881)(22,  -0.09689269214868546)(23,  -0.10385064780712128)(24, -0.09194887429475784)(25,  -0.07049514353275299)(26,  -0.08572334051132202)(27,  -0.14785674214363098)(28, -0.21511705219745636)(29,  -0.2652265727519989)(30,  -0.28198063373565674)(31,  -0.25845178961753845)(32, -0.22051864862442017)(33,  -0.20809803903102875)(34,  -0.18731571733951569)(35,  -0.18151743710041046)(36, -0.1867053508758545)(37,  -0.17651259899139404)(38,  -0.161528542637825)(39,  -0.14791779220104218)(40, -0.12182539701461792)(41,  -0.1132805347442627)(42,  -0.11840745806694031)(43,  -0.12316816300153732)(44, -0.11834640055894852)(45,  -0.09377995878458023)(46,  -0.05368007719516754)(47,  -0.029083071276545525)(48, -0.019348010420799255)(49,  -0.04159518703818321)}; \addlegendentry{$\delta=-1$};
    \end{axis}
\end{tikzpicture}
    \vspace{-0.5\baselineskip}
  \caption{Plot of original speech signal $\vx [n]$ (black) and approximated time-shifted signals $\Phi(\vx,\delta\vphi_\mathrm{ref})$ by PhaseAug with $\delta=+1$ (red), $+0.5$ (orange), $-0.5$ (green), and $-1$ (blue).}
  \label{fig:plot_shifted}
    %  \vspace{-0.25\baselineskip}
\end{figure}

From (\ref{eq_time_shift}) and (\ref{eq_phaseaug}), time-shifted signal $\vx[n-\delta]$, where $\delta \in \mathbb{Z}$, can be approximated by PhaseAug as:
\begin{equation}
    \vx[n-\delta] \approx
    \Phi \left(\vx[n],
    - \delta \vphi_\mathrm{ref}
    \right)
    , \label{eq_phaseaug_timeshift}
\end{equation} 
where $\vphi_\mathrm{ref} = 2\pi/N \cdot \begin{bmatrix} 0 &1&2& \ldots& N/2 \end{bmatrix}$ is a reference phase-rotation vector.
With PhaseAug, time-shift can be approximated as a continuous and differentiable operation by controlling the time-shift value $\delta \in \mathbb{R}$, 
thus we can generate approximated non-integer time-shifted signal by calculating $\Phi(\vx, -\delta \vphi_\mathrm{ref})$. 
In Fig. \ref{fig:plot_shifted}, we demonstrate that PhaseAug can approximate continuous and differentiable time-shift. 

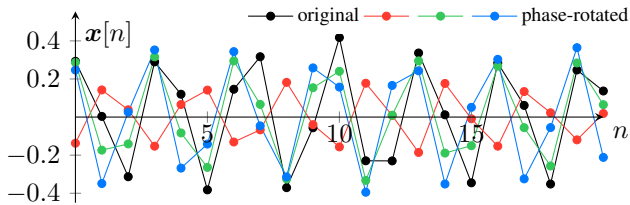
\begin{figure}[h]
  \centering
  \begin{tikzpicture}
    \usepgfplotslibrary{polar}
    \usepgflibrary{shapes.geometric}
    \pgfplotsset{mystyle/.append style={axis x line=middle,
    axis y line=
           middle,
    xlabel={$n$}, ylabel={$\vx[n]$},
    xticklabel={$\pgfmathprintnumber{\tick}$},
    yticklabel={$\pgfmathprintnumber{\tick}$},
    set layers,% using layers
    axis on top,
    },
    }
    \begin{axis}[mystyle,width = \linewidth,height = 0.48\linewidth,
    legend style={draw=none, fill opacity=1., text opacity = 1,row sep=0pt,nodes={scale=0.75, transform shape}, at={(0.31, 0.975)},anchor=west},
    legend columns=4,% transpose legend,    
    xtick={21,26,31,36},
    xticklabels={$0$,$5$,$10$,$15$},
    % extra x ticks={26, 36},
    % extra x tick labels={$5$, $15$},
    % extra x tick style={
    xticklabel style={yshift=2.5},
    x label style={anchor=north west},
    ytick={-0.4,-0.2,0,0.2,0.4}, 
    xmin=21, xmax=41, ymin=-0.45, ymax=0.555]
    %\addlegend{original,$\delta=-1$,$\delta=-0.5$,$\delta=+0.5$,$\delta=+1$}
        \addplot[mark=*, color = black,mark options  = {scale = 0.75}] coordinates {
(0,-0.085174560546875)(1,-0.084686279296875)(2,0.12786865234375)(3,0.028594970703125)(4,-0.162933349609375)(5,0.139251708984375)(6,0.06378173828125)(7,-0.1842041015625)(8,0.104949951171875)(9,0.1107177734375)(10,-0.199554443359375)(11,0.06622314453125)(12,0.194793701171875)(13,-0.2259521484375)(14,-0.040924072265625)(15,0.320343017578125)(16,-0.18316650390625)(17,-0.18707275390625)(18,0.341064453125)(19,-0.07513427734375)(20,-0.24493408203125)(21,0.2928466796875)(22,0.0035400390625)(23,-0.313323974609375)(24,0.2896728515625)(25,0.119873046875)(26,-0.382293701171875)(27,0.146026611328125)(28,0.316864013671875)(29,-0.370697021484375)(30,-0.0560302734375)(31,0.417572021484375)(32,-0.22955322265625)(33,-0.2301025390625)(34,0.33642578125)(35,0.01190185546875)(36,-0.34521484375)(37,0.28411865234375)(38,0.0611572265625)(39,-0.3516845703125)(40,0.247222900390625)(41,0.1368408203125)(42,-0.349853515625)(43,0.090362548828125)(44,0.265350341796875)(45,-0.28533935546875)(46,-0.06787109375)(47,0.29962158203125)(48,-0.130859375)(49,-0.216156005859375)
}; \addlegendentry{original};
        \addplot[mark=*, color = ired, mark options  = {scale = 0.75}] coordinates {
(0,0.10245358198881149)(1,0.01010231114923954)(2,-0.10647964477539062)(3,0.06048067286610603)(4,0.1390298306941986)(5,-0.21024593710899353)(6,0.06387139111757278)(7,0.17675426602363586)(8,-0.20922410488128662)(9,-0.013089914806187153)(10,0.2607243061065674)(11,-0.16295596957206726)(12,-0.15449602901935577)(13,0.3057463467121124)(14,-0.10986075550317764)(15,-0.1480722725391388)(16,0.2701992392539978)(17,-0.11673551797866821)(18,-0.13201840221881866)(19,0.22721458971500397)(20,-0.04545186460018158)(21,-0.1373099535703659)(22,0.14238791167736053)(23,0.038220737129449844)(24,-0.15288302302360535)(25,0.0659436583518982)(26,0.14159218966960907)(27,-0.13088953495025635)(28,-0.06717058271169662)(29,0.18221580982208252)(30,-0.040054306387901306)(31,-0.15668702125549316)(32,0.17764239013195038)(33,0.008748254738748074)(34,-0.18574923276901245)(35,0.1764863133430481)(36,-0.008909855037927628)(37,-0.1527818888425827)(38,0.13393409550189972)(39,0.022537104785442352)(40,-0.11961080878973007)(41,0.018765173852443695)(42,0.08519729971885681)(43,-0.041801583021879196)(44,-0.11801955103874207)(45,0.13626031577587128)(46,0.017254464328289032)(47,-0.18313056230545044)(48,0.13570243120193481)(49,0.027876293286681175)
}; \addlegendentry{};
\addplot[mark=*, color = igreen, mark options  = {scale = 0.75}] coordinates {
(0,-0.21887663006782532)(1,-0.04428372159600258)(2,0.302731990814209)(3,-0.15586885809898376)(4,-0.17657753825187683)(5,0.2774207592010498)(6,-0.007639870047569275)(7,-0.2318311482667923)(8,0.1532948762178421)(9,0.15197579562664032)(10,-0.268274188041687)(11,0.09003226459026337)(12,0.2131628692150116)(13,-0.2842748165130615)(14,0.03805205598473549)(15,0.3072410821914673)(16,-0.28841716051101685)(17,-0.054951127618551254)(18,0.3440558910369873)(19,-0.24864232540130615)(20,-0.05882622301578522)(21,0.28685462474823)(22,-0.17368629574775696)(23,-0.14100484549999237)(24,0.3162432909011841)(25,-0.08325491845607758)(26,-0.26422828435897827)(27,0.2946506142616272)(28,0.06667789071798325)(29,-0.3249754011631012)(30,0.15410344302654266)(31,0.2402176707983017)(32,-0.33239999413490295)(33,0.00959612987935543)(34,0.29501211643218994)(35,-0.18902097642421722)(36,-0.15032033622264862)(37,0.26721084117889404)(38,-0.055632371455430984)(39,-0.2568991482257843)(40,0.2841241955757141)(41,0.06505659967660904)(42,-0.4143202006816864)(43,0.259037584066391)(44,0.2190515697002411)(45,-0.47238826751708984)(46,0.16869981586933136)(47,0.31677040457725525)(48,-0.4086189568042755)(49,-0.05335158482193947)}; \addlegendentry{};
    \addplot[mark=*, color = iblue, mark options  = {scale = 0.75}] coordinates {
(0,-0.2050735205411911)(1,0.16135913133621216)(2,0.09463685005903244)(3,-0.22091513872146606)(4,0.10589614510536194)(5,0.16470718383789062)(6,-0.25652584433555603)(7,0.0677192285656929)(8,0.21480800211429596)(9,-0.2099117636680603)(10,-0.060984328389167786)(11,0.28173863887786865)(12,-0.1352887749671936)(13,-0.1774081438779831)(14,0.25905323028564453)(15,0.04314021021127701)(16,-0.2868938744068146)(17,0.1856476068496704)(18,0.16737210750579834)(19,-0.3394569158554077)(20,0.1478157937526703)(21,0.2469174861907959)(22,-0.34905245900154114)(23,0.02614239975810051)(24,0.35245436429977417)(25,-0.267909437417984)(26,-0.1418149620294571)(27,0.3436351418495178)(28,-0.046324074268341064)(29,-0.31405508518218994)(30,0.2582646310329437)(31,0.15742269158363342)(32,-0.39467838406562805)(33,0.16618967056274414)(34,0.24283407628536224)(35,-0.3514920175075531)(36,0.05095010995864868)(37,0.3031793534755707)(38,-0.32449084520339966)(39,-0.055107295513153076)(40,0.3644624650478363)(41,-0.21171289682388306)(42,-0.19602228701114655)(43,0.28485962748527527)(44,-0.061061251908540726)(45,-0.14715859293937683)(46,0.12724469602108002)(47,-0.01951770856976509)(48,-0.07378894835710526)(49,0.05327378585934639)
}; \addlegendentry{phase-rotated};

    \end{axis}
\end{tikzpicture}
    \vspace{-2.\baselineskip}
  \caption{Plot of original speech signal $\vx[n]$ (black) and phase-rotated signals $\Phi(\vx,\vphi)$ by PhaseAug with augmentation policy with fixed mean time-shift as $\delta=0$ (red, green, blue).}
  \label{fig:plot_phaseshifted}
%   \vspace{-.5\baselineskip}
\end{figure}

For random augmentation, we first attempted to sample phase-rotation $\vmu$ from $\vmu\sim \mathcal{N}(\delta \vphi_\mathrm{ref}, \sigma^2 \mI)$ and augmented signal as $\vx^\mathrm{aug}=\Phi(\vx, \vmu)$, but rotating phases by the same variance affects each frequency bin differently.
Since $\vmu \odot \vphi_\mathrm{ref}^{-1}$ is shifted time for each frequency bin by (\ref{eq_phaseaug_timeshift}), 
where $\vphi_\mathrm{ref}^{-1}=\begin{bmatrix} 0 & 1/\phi_\mathrm{ref}[1] & 1/\phi_\mathrm{ref}[2]& \ldots& 1/\phi_\mathrm{ref}[N/2] \end{bmatrix}$ and $\odot$ is element-wise multiplication, low-frequency bins are more shifted in time than high-frequency bins.
To calibrate variances of frequency bins in the time domain, we consider $\vmu \sim \mathcal{N}(\delta \bm{1}, \sigma^2 \mI)$ as a time-shift for each frequency bin instead of phase-rotation by calculating $\vx^\mathrm{aug}=\Phi(\vx, \vmu \odot \vphi_\mathrm{ref})$.
In addition, we observed that $\vx^\mathrm{aug}$ has noticeable distortion with $\vmu \sim \mathcal{N}(\delta \bm{1}, \sigma^2 \mI)$.
Since Karras \etal{} \cite{stylegan2ada} state that leaking of augmentation has occurred and we also noticed it with the setting above, we attempted to reduce the distortion.
We empirically found that a large time-shift difference between adjacent frequency bins occurs in distortion and high-frequency elements in $\vmu$ cause a large difference between adjacent frequency bins. 
To reduce perceptible distortion by removing high-frequency components,
we apply a low-pass filter (LPF) to calculate filtered time-shift $\vmu_l=\mathrm{LPF}(\vmu)$, specifically the Kaiser-windowed sinc filter. 
Moreover, we set $\vmu$'s variance $\sigma^2>5.2$ to keep the variance of $\vmu_l$ greater than $0.5$ since LPF in our setup reduces variance by $90.3\%$.
Fig. \ref{fig:plot_phaseshifted} illustrates the original signal and its random phase-rotated versions which are indistinguishable by hearing. 
    
% \vspace{-.5\baselineskip}
\subsection{Augmentation Policy}
% \vspace{-.1\baselineskip}
For random augmentation, we apply the following Algorithm \ref{alg:phaseaug_policy} to real and generated signals.
\begin{algorithm}[H]
    \algrenewcommand\algorithmicindent{1.0em}
  \caption{PhaseAug augmentation policy} \label{alg:phaseaug_policy}
  \hspace*{\algorithmicindent} \textbf{Input}: signal $\vx$ \\
  \hspace*{\algorithmicindent} \textbf{Output}: augmented signal $\vx^\mathrm{aug}$ 
  \begin{algorithmic}[1]
   
      \State Sample $\delta \sim \mathcal{U}(-\delta_\mathrm{max},\delta_\mathrm{max})$
      \State Sample $\vmu \sim \mathcal{N}(\delta\1,\sigma^2 \mI)$
      \State Compute $\vmu_l \leftarrow \mathrm{LPF}(\vmu)$
      \State Compute $\vphi \leftarrow \vmu_l \odot \vphi_\mathrm{ref}$  
      \State Compute $\vx^\mathrm{aug} \leftarrow \Phi(\vx, \vphi)$
  \end{algorithmic}
\end{algorithm}
% \vspace{-.2\baselineskip}

\subsection{Differentiable Augmentation for GAN-based Vocoder}
% \vspace{-.1\baselineskip}
Since GAN-based vocoders are adopting feature matching loss, we apply PhaseAug with identical $\vphi$ to real and generated samples.
For paired raw waveform $\vx$ and mel-spectrogram $\ms$, the discriminators take the inputs as $\Phi(\vx, \vphi)$ and $\Phi(G(\ms), \vphi)$, where $G$ is the generator model.
For a sample, we employ the same phase-rotation $\vphi$ to all discriminators due to computational efficiency.
In addition, we sample different $\vphi$ between discriminators update and generator update for providing more points of view of waveforms.
Mel-spectrogram loss is calculated between $\vx$ and $G(\ms)$, not between $\Phi(\vx, \vphi)$ and $\Phi(G(\ms), \vphi)$.

\subsection{Implementation Details}
We use one-sided STFT with $N=1024$ and $d=256$.
We employ Kaiser-windowed sinc kernel with kernel size $128$, cut-off frequency of $0.05$, and transition band half-width $0.012$ for the LPF, and $\mathrm{LPF}$ in Algorithm \ref{alg:phaseaug_policy} denotes $1$D convolution with the specified kernel. %\footnote{We adopted filter from \url{https://github.com/junjun3518/alias-free-torch}.}.
We sample $B$ different $\vphi$ per step, where $B$ is the batch size during training.
We calculate complex tensor as real-valued length $2$ vector and multiplying $e^{j\phi[k]}$ is replaced by multiplying rotation matrix $\big(\begin{smallmatrix} \cos{\phi[k]} &-\sin{\phi[k]} \\ \sin{\phi[k]} & \cos{\phi[k]}
\end{smallmatrix}\big)$.
Since time-shift with small enough $\delta$ does not cause perceivable artifacts, we set the maximum value for $\delta$ to $\delta_\mathrm{max}=2$ which is small enough than $N$ to maintain the mel-spectrogram condition.
In the same manner, we set $\sigma^2=6$ to get $\vmu_l$ with variance approximately $0.58$.
PhaseAug is applied with $100\%$ augmentation probability with identical augmentation strength.
We do not employ ADA \cite{stylegan2ada}, since the augmentation probability converged to $100\%$ in every setup of our initial experiments.

\section{Experiments}
\subsection{Baselines}
To evaluate PhaseAug, we train HiFi-GAN %\footnote{Official implementation \url{https://github.com/jik876/hifi-gan}}
\cite{hifigan}, which is a state-of-the-art GAN-based neural vocoder model.
In addition, we simply replace the HiFi-GAN generator to the MelGAN generator %\footnote{Official implementation \url{https://github.com/descriptinc/melgan-neurips}}
\cite{melgan} without changing other setups such as mel-spectrogram loss, loss coefficients, MPD, \etc for generator architecture ablation.
% To compare vocoder models with and without PhaseAug, we trained MelGAN\footnote{Official implementation \url{https://github.com/descriptinc/melgan-neurips}} \cite{melgan} 
%  and HiFi-GAN\footnote{Official implementation \url{https://github.com/jik876/hifi-gan}} \cite{hifigan} to evaluate PhaseAug.
During the training, we follow all configurations, with exception of PhaseAug and size of the dataset, as the official HiFi-GAN $V1$ configurations.
All models are trained from scratch on a V$100$ GPU.

\subsection{Dataset}
We trained all models on LJSpeech \cite{ljspeech}, which contains $24$ hours of single-speaker speeches, and divided the training set and the validation set as official HiFi-GAN configurations.
Since it contains sufficient amount of files to train vocoder models, we conduct training with $100\%$, $10\%$, and $1\%$ of the dataset to evaluate the effect of PhaseAug on various sizes of the dataset including limited data conditions.
We take each percentage depending on the number of files and not on the total playtime due to the different playtime of each file.
$100\%$ and $10\%$ models are trained until $2.5$M steps and $1\%$ models are stopped before their validation mel-spectrogram errors diverge ($\leq 100$k steps).
For limited data size conditions, we choose a model with the minimum validation mel-spectrogram mean average error. In addition, we report means and standard deviations over $3$ trainings for evaluation metrics of $1\%$ models, because of there high variance.

\tableresults{}

\subsection{Evaluation Metrics}
\vspace{-0.74\baselineskip}
%need to paraphrase?
To evaluate our methods, we measure mel-spectrogram mean average error (MAE),
multi-resolution STFT (M-STFT) %\footnote{Unofficial implementation \url{https://github.com/csteinmetz1/auraloss}}
\cite{parallelwavegan},
$16000$ Hz wide-band perceptual evaluation of speech quality (PESQ) %\footnote{Unofficial implementation \\ \url{https://github.com/ludlows/python-pesq}}
\cite{pesq}, 
mel-cepstral distortion (MCD) %\footnote{Unofficial implementation \url{https://github.com/ttslr/python-MCD}} 
\cite{mcd},
periodicity error \cite{cargan}, 
and F$1$ score of voiced/unvoiced classification (V/UV F$1$) %\footnote{Official implementation \url{https://github.com/descriptinc/cargan}}
\cite{cargan} as objective metrics.
% PESQ https://github.com/ludlows/python-pesq
% MCD https://github.sscom/ttslr/python-MCD.
%periodicity / V/UV CARGAN
We conduct subjective evaluation using mean opinion score (MOS) and pairwise preference test.
Each of MOS is performed with $150$ samples and $750$ ratings, a with $95\%$ confidence interval (CI).
In the pairwise preference test, each pair is evaluated using $150$ samples and $1500$ votes.
Participants are asked to select a preferred sample between two models, with or without PhaseAug, or a neutral option.

\vspace{-0.74\baselineskip}
\section{Results} \label{result}
\vspace{-0.74\baselineskip}
Table \ref{tab:result} denotes the results from the models trained with or without PhaseAug.
Without some cases of MCD and V/UV F1, our method improves the evaluation metrics for HiFi-GAN and MelGAN generators.
Especially, PhaseAug improves periodicity (average $3\%$ decreased), which is the most important metric related to the artifacts. 
In addition, models with $10\%$ dataset and PhaseAug show better MAE than models with $100\%$ dataset and without PhaseAug.
There are no statistically significant differences between MOS tests.
MOS of augmented ground truth signals also do not have significant difference between ground truth signals.
However, models with PhaseAug show higher preferences with significant differences without MelGAN $1\%$ case, since $1\%$ of dataset is not enough to train MelGAN.

\vspace{-0.74\baselineskip}
\section{\small{Discussion, Limitation, and Future Work}}
\vspace{-0.74\baselineskip}
In this work, we propose PhaseAug, the first differentiable augmentation for speech synthesis to overcome the discriminator overfitting and the one-to-one paired data problem.
Our method rotates the phases of speech to simulate one-to-many mapping for vocoder discriminators. 
With PhaseAug, we could enhance the value of evaluation metrics without modifying network architectures. 
Since PhaseAug is also effective in limited dataset cases, we expect that our method could be applicable during fine-tuning GAN-based vocoder with a small dataset.

While Zhao \etal{} \cite{diffaug} and Karras \etal{} \cite{stylegan2ada} report that DiffAugment could prevent discriminator overfitting,
we observe that MPD and MSD show similar accuracy patterns with or without PhaseAug.
We expect that this is caused by the difference between unconditional and conditional generative models.
While accuracy patterns are similar, mean of MPD outputs (closer to $0.5$) are shown that MPD provides more useful information to the generator with our method.
In addition, leaking of PhaseAug is negligible, since MOS of augmented signals do not have significant difference with ground truth.

We compare our method with several objective evaluation metrics.
However, these metrics have limitations as they measure the difference between the generated signals and the ground truth signals in a one-to-many mapping problem.
Moreover, periodicity artifact is observed in short moments, these metrics are not appropriate for evaluating it. 
Listeners focus on samples' global naturalness, not temporal artifact, so MOS is not an appropriate metric for it. However, pairwise test shows significant differences.
We also tested our method to VITS \cite{vits}, a GAN-based end-to-end text-to-speech (E2E TTS) model. 
However measuring objective evaluation metrics was difficult as E2E TTS is a more complex one-to-many problem than neural vocoding.

Since PhaseAug could approximate differentiable time-shift, we could train a shift-equivariant speech synthesis model by applying it to hidden latents and output of model without filtered nonlinearity \cite{bigvgan,aliasfreegan}.
In addition, PhaseAug is also applicable as data augmentation to other speech synthesis models such as neural audio upsampling \cite{nuwave, nuwave2}, while we focus on applying it as a DiffAugment for GAN-based vocoders.

\vspace{-0.71\baselineskip}
\section{Acknowledgment}
\vspace{-0.71\baselineskip}
The authors would like to thank Minho Kim of MINDsLab, Jinwoo Kim and Hyeonuk Nam from KAIST, Dongho Choi and Kang-wook Kim from SNU for valuable discussions.

% \vfill\pagebreak

% References should be produced using the bibtex program from suitable
% BiBTeX files (here: strings, refs, manuals). The IEEEbib.bst bibliography
% style file from IEEE produces unsorted bibliography list.
% -------------------------------------------------------------------------

\bibliographystyle{IEEEbib}
\bibliography{icassp_phaseaug}

\end{document}